\documentclass[aps,showpacs]{revtex4}
\usepackage{bm}
\usepackage{graphicx}
\usepackage{amsbsy}
\usepackage{amsmath}
\usepackage{amsfonts}

\newcommand{\bra}[1]{\langle{#1}|}
\newcommand{\ket}[1]{|{#1}\rangle}
\newcommand{\nn}{\nonumber}
\newcommand{\dg}{^\dagger}
\newcommand{\acos}{\text{acos}}
\newcommand{\per}{\text{per}}

\begin{document}

\title{Post-processing with linear optics for improving the quality of
single-photon sources}

\author{Dominic W. Berry$^1$, Stefan Scheel$^2$, Casey R. Myers$^4$, \\
Barry C. Sanders$^{1,3}$, Peter L. Knight$^2$ and Raymond Laflamme$^{4,5}$} 
\affiliation{${}^1$ Australian Centre for Quantum Computer Technology,
Department of Physics, Macquarie University, Sydney, New South Wales 2109,
Australia \\
$^2$ Quantum Optics and Laser Science, Blackett Laboratory, 
Imperial College London, Prince Consort Road, London SW7 2BW, UK \\ 
$^3$ Quantum Information Science Group, Department of Physics and
Astronomy, University of Calgary, Alberta T2N 1N4, Canada \\
$^4$ Institute for Quantum Computing, University of Waterloo, ON, N2L 3G1,
Canada \\
$^5$ Perimeter Institute for Theoretical Physics, 35 King Street N., Waterloo,
ON, N2J 2W9, Canada}

\begin{abstract}
Triggered single-photon sources produce the vacuum state with non-negligible
probability, but produce a much smaller multiphoton component. It is therefore
reasonable to approximate the output of these photon sources as a mixture of the
vacuum and single-photon states. We show that it is impossible to increase the
probability for a single photon using linear optics and photodetection on fewer
than four modes. This impossibility is due to the incoherence of the inputs; if
the inputs were pure-state superpositions, it would be possible to obtain a
perfect single-photon output. In the more general case, a chain of beam
splitters can be used to increase the probability for a single photon, but at
the expense of adding an additional multiphoton component. This improvement is
robust against detector inefficiencies, but is degraded by dark counts or
multiphoton components in the input.
\end{abstract}

\date{\today}

\pacs{03.67.-a, 42.50.Dv}

\maketitle

\section{Introduction}
One of the most promising methods for quantum information processing is linear
optics and photodetection. Linear optics and photodetection may be used for
provably secure quantum communication \cite{BB84}, as well as quantum
computation \cite{LOQC}. An important requirement for these schemes is the
ability to produce a single photon on demand \cite{key,LOQC}, yet generating
high-fidelity single-photon states is challenging. The traditional method for
generating single photons involves photodetection on one output mode from a
non-degenerate parametric down-conversion process to post-select a single photon
in the correlated mode \cite{down,fid}. This method has the drawback that the
time of the photon emission is not controlled. More recently, triggered photon
sources have been developed, including molecules \cite{mol}, quantum wells
\cite{well}, colour centers \cite{col}, ions \cite{ion} and quantum dots
\cite{dot}. These sources have a significant vacuum contribution, but the
multiphoton contribution may be made very small \cite{twophot}.

For the majority of this study, we approximate these sources by taking the
multiphoton probability to be zero. That is, we consider an idealised
single-mode single-photon source, which may be represented by the density
operator
\begin{equation} \label{eq:initialstate}
\hat{\rho}_p = (1-p) \ket 0 \bra 0 + p \ket 1 \bra 1 .
\end{equation}
Here $p$ is the probability for a single photon, and is also called the
efficiency. Increasing the efficiency is important because many quantum optics
experiments, especially those concerned with linear optical quantum computation,
require high efficiency sources. We also derive results for states with a
coherent superposition of the vacuum and a single photon; however, this form of
state is not produced by single-photon sources.

Much effort is directed towards improving sources, but here we pose the
question as to whether it is possible to perform post-processing to obtain
higher efficiency. Ideally this post-processing should also maintain a zero
multiphoton contribution (though a very small multiphoton contribution would
also be acceptable). A promising method of post-processing is linear optics
and photodetection. As mentioned above, linear optics and photodetection can
be used to perform quantum computation \cite{LOQC}, and optical controlled-NOT
gates have recently been demonstrated \cite{CNOT}; however, there are also no-go
theorems for linear optics \cite{nogo}.

In a recent publication \cite{BSSK} we investigated the possibility of improving
the single-photon efficiency $p$ with linear optical elements and
photodetection. Here we present a number of new results that clarify the
limitations inherent in this method, as well as reviewing the results presented
in \cite{BSSK}. In particular, we analyse the effects of various experimental
limitations, present a scheme that gives perfect results for inputs with a
coherent superposition of zero and one photon, and show connections between some
difficult unsolved problems.

We begin by showing that it is not possible to obtain an improvement for the
simple case of a beam splitter in Sec.\ \ref{Sec:split}.
In Sec.\ \ref{Sec:coherent} we show that, if we allow a coherent superposition
of zero and one photon, rather than an incoherent mixture, it is possible to
obtain a perfect single-photon state. We then proceed to the case of a multimode
interferometer with incoherent inputs in Sec.\ \ref{Sec:general}. In
Secs.\ \ref{Sec:limit} and \ref{Sec:method} we give an expanded discussion of
the limit on the improvement that it is possible to obtain, and the method to
obtain an improvement. Sec.\ \ref{Sec:exp} gives a detailed discussion of the
impact of various experimental problems on this method. We give further
discussion of no-go theorems for post-processing in Sec.\ \ref{Sec:nogo}. Then,
in Sec.\ \ref{Sec:conn}, we show that there are deep connections between the
unsolved problems for post-processing. Lastly we give a discussion of how the
theory is changed by allowing multiphoton contributions in the inputs in
Sec.\ \ref{Sec:multi}, and we conclude in Sec.\ \ref{Sec:conc}.

\section{Beam splitter}
\label{Sec:split}
To begin, we consider the simplest case of two copies
of the quantum state (\ref{eq:initialstate}) combined on a single beam
splitter. The initial state can be written in the form
\begin{equation}
\hat{\rho}_{\text{in}}^{(2)} = (1-p_1)(1-p_2) \ket{00}\bra{00} +p_1(1-p_2)
\ket{10}\bra{10}+p_2(1-p_1) \ket{01}\bra{01} +p_1p_2\ket{11}\bra{11} .
\end{equation}
Each photonic mode operator gets transformed by the beam splitter in
the following way:
\begin{equation}
\label{splitter}
\hat{a}_1\dg\mapsto\Lambda_{11}\hat{a}_1\dg+\Lambda_{21}\hat{a}_2\dg , \qquad
\hat{a}_2\dg\mapsto\Lambda_{12}\hat{a}_1\dg+\Lambda_{22}\hat{a}_2\dg ,
\end{equation}
where $\bm{\Lambda}$ is a $2\times 2$ unitary matrix \cite{VogelWelsch}. We have
two options, to project onto the vacuum or onto the single-photon Fock state
(projecting onto two photons results in vacuum output in the unmeasured mode).
After vacuum projection in mode 2, we end up with an unnormalised state of the
form
\begin{equation}
\hat{\rho}_{\text{out}}^{(2)} \propto \ket{0}\bra{0}
+\left( \frac{p_1}{1-p_1}|\Lambda_{11}|^2 + \frac{p_2}{1-p_2}|\Lambda_{12}|^2
\right) \ket{1}\bra{1} +\text{2-photon term} .
\end{equation}
Thus the ratio between the probabilities for one and zero photons is just a
weighted average of $p_1/(1-p_1)$ and $p_2/(1-p_2)$, and cannot exceed either of
these. That is, it is not possible to improve the ratio between the
probabilities for obtaining one and zero photons. This automatically implies
that it is not possible to improve the absolute probability of obtaining one
photon.

In the case of a single-photon detection, the resulting unnormalised state is
\begin{equation}
\hat{\rho}_{\text{out}}^{(2)} \propto \left( \frac{1-p_1}{p_1}
|\Lambda_{22}|^2 +\frac{1-p_2}{p_2} |\Lambda_{21}|^2 \right)
\ket{0}\bra{0} +|\per\bm{\Lambda}|^2 \ket{1}\bra{1} ,
\end{equation}
where $\per\bm{\Lambda}$ is the permanent \cite{perm} of the beam splitter
matrix $\bm{\Lambda}$. Since the absolute value of the permanent of a unitary
matrix is bounded from above by unity, and the term in brackets is a weighted
sum of terms $(1-p_i)/p_i$, we do not find any improvement for this case either.
Hence there is no improvement in the probability for a single photon if zero,
one or two photons are detected. These results demonstrate that for mixed-state
inputs it is impossible to obtain an improvement in the single-photon
probability using a beam splitter.

\section{Pure-state inputs}
\label{Sec:coherent}
It is possible to obtain an improvement using a beam splitter if the inputs are
in pure-state superpositions of zero and one photon, instead of incoherent
mixtures. Consider two input modes that are each in the state
$\alpha\ket{0}+\beta\ket{1}$. The initial state may be written as
\begin{equation}
\ket{\psi}_{\text{in}}^{(2)} = \left[\alpha^2+\alpha\beta (a_1\dg + a_2\dg) +
\beta^2 a_1\dg a_2\dg \right] \ket{00} .
\end{equation}
Applying the beam splitter transformation \eqref{splitter} gives
\begin{equation}
\ket{\psi}_{\text{trans}}^{(2)} = \left\{\alpha^2+\alpha\beta \left[
(\Lambda_{11}+\Lambda_{12}) a_1\dg + (\Lambda_{21}+\Lambda_{22}) a_2\dg \right]
+ \beta^2 \left[\Lambda_{11}\Lambda_{12}(a_1\dg)^2 + \Lambda_{21}\Lambda_{22}
(a_2\dg)^2 + (\per\bm{\Lambda}) a_1\dg a_2\dg \right]\right\} \ket{00} .
\end{equation}
Conditioning on detection of zero photons in mode 2 gives the output state
\begin{equation}
\ket{\psi}_{\text{out}}^{(2)} \propto \alpha^2\ket{0} + \alpha\beta
(\Lambda_{11} + \Lambda_{12})\ket{1} + \sqrt{2}\beta^2\Lambda_{11}\Lambda_{12}
\ket{2}.
\end{equation}
It is easily seen that this output state may have a higher probability for a
single photon. For example, if the initial state is close to the vacuum state
(i.e.\ $\alpha\gg\beta$), then an improvement by a factor of two may be obtained
by using $\Lambda_{11}=\Lambda_{12}=1/\sqrt{2}$.

In fact, it is possible to further process this output state to obtain a perfect
single-photon state. If we combine this state with mode 3, which is also assumed
to be prepared in state $\alpha\ket{0}+\beta\ket{1}$, then the total state may
be represented as
\begin{equation}
\ket{\psi}_{\text{in}}^{(3)} \propto \left[ \alpha^2 + \alpha\beta (\Lambda_{11}
+\Lambda_{12})a_1\dg + \beta^2\Lambda_{11}\Lambda_{12}(a_1\dg)^2 \right]
\left( \alpha + \beta a_3\dg \right) \ket{00}.
\end{equation}
Applying the beam splitter transformation \eqref{splitter} (except using a prime
to distinguish this beam splitter from the previous one) gives
\begin{equation}
\ket{\psi}_{\text{trans}}^{(3)} \propto \left[ \alpha^2 + \alpha\beta
(\Lambda_{11} + \Lambda_{12})(\Lambda'_{11} \hat{a}_1\dg +\Lambda'_{31}
\hat{a}_3\dg) + \beta^2\Lambda_{11}\Lambda_{12}(\Lambda'_{11} \hat{a}_1\dg +
\Lambda'_{31} \hat{a}_3\dg)^2 \right] \left[ \alpha + \beta (\Lambda'_{13}
\hat{a}_1\dg +\Lambda'_{33} \hat{a}_3\dg) \right] \ket{00}.
\end{equation}
Conditioning on detection of two photons gives
\begin{equation}
\ket{\psi}_{\text{out}}^{(3)} \propto \sqrt{2}\beta^2 \Lambda'_{31} \left\{
\alpha\left[(\Lambda_{11}+\Lambda_{12})\Lambda'_{33}+\Lambda_{11}\Lambda_{12}
\Lambda'_{31} \right]\ket{0}+\beta \Lambda_{11}\Lambda_{12}\left( 2\Lambda'_{11}
\Lambda'_{33}+\Lambda'_{31}\Lambda'_{13}\right)\ket{1} \right\}
\end{equation}

To make this equation more clear, we use
\begin{equation}
\bm{\Lambda} = \left[ \begin{array}{*{20}c} e^{i\phi}\cos{\theta} & -\sin
{\theta} \\ \sin\theta & e^{-i\phi}\cos\theta  \\ \end{array} \right], \qquad
\bm{\Lambda}' = \left[ \begin{array}{*{20}c} e^{i\phi'}\cos\theta'  & -\sin
\theta' \\ \sin\theta' & e^{-i\phi'}\cos\theta'  \\ \end{array} \right].
\end{equation}
The condition that the output state is a pure single-photon state then becomes
\begin{equation}
(\cos\theta -e^{-i\phi}\sin\theta)e^{-i\phi'}\cos\theta'-\sin\theta\cos\theta
\sin\theta' =0.
\end{equation}
This equation may be satisfied by taking $\theta'$ and $\phi'$ to be
\begin{equation}
\label{thprim}
\theta'=\arctan \left( \frac{|\cos\theta-e^{-i\phi}\sin\theta|}{\sin\theta\cos
\theta} \right), \qquad \phi' = \arg(\cos\theta -e^{-i\phi}\sin\theta).
\end{equation}
That is, regardless of the characteristics of the initial beam splitter
(provided $\sin\theta$ and $\cos\theta$ are nonzero), it is possible to obtain a
perfect single-photon output.

\begin{figure}[t]
\centerline{\includegraphics[width=8cm]{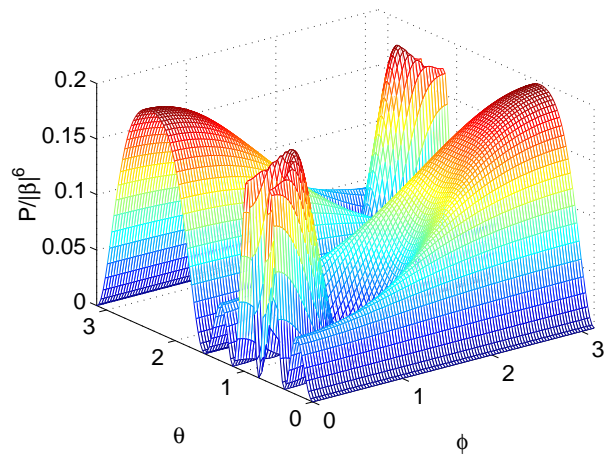}}
\caption{\label{fig:probs} The probability of obtaining the desired detection
results as a function of the beam splitter parameters $\theta$ and $\phi$ for
the first beam splitter.}
\end{figure}

Another issue is the probability for obtaining the desired pattern of detection
results. Using the unnormalised expression for $\ket{\psi}_{\text{out}}^{(3)}$
above, this probability is given by
\begin{align}
P &= 2|\beta|^6 \sin^2\theta\cos^2\theta\sin^2\theta' (2\cos^2\theta'-\sin^2
\theta')^2 \nn \\
&= |\beta|^6 \frac{\frac 12 \sin^2 2\theta (1-\cos\phi \sin 2\theta) \left(
\frac 12 \sin^2 2\theta -1+\cos\phi\sin 2\theta \right)^2}{\left( \frac 14
\sin^2 2\theta +1-\cos\phi\sin 2\theta \right)^3}.
\end{align}
In the second line we have used the expression \eqref{thprim} for $\theta'$.
This probability is plotted in Fig.\ \ref{fig:probs} for the range $0$ to $\pi$
in $\theta$ and $\phi$ ($P$ is periodic with period $\pi$ in these variables).
There are four maxima in this range, for $(\theta,\phi)=(\pi/4,\pi)$,
$(\pi/4,\acos(13/14))$, $(3\pi/4,0)$ and $(3\pi/4,\acos(-13/14))$. The exact
values of $\phi$ of $\acos(13/14)$ and $\acos(-13/14)$ are not obvious from the
plot but are straightforward to obtain analytically. The two maxima
$(\pi/4,\pi)$ and $(3\pi/4,0)$ correspond to the same beam splitter, so there
are only three maxima that correspond to distinct beam splitters. Each of these
maxima is exactly the same height,
\begin{equation}
P_{\text{max}} = 16|\beta|^6/81.
\end{equation}

One factor that distinguishes the maxima is the sensitivity to the parameters.
Clearly the maxima at $(\pi/4,\acos(13/14))$ and $(3\pi/4,\acos(-13/14))$ are
far more sensitive to the values of $\theta$ and $\phi$, and it is therefore
better to use the beam splitter corresponding to $(\pi/4,\pi)$ and $(3\pi/4,0)$.
For the second beam splitter, the appropriate parameters are $\theta'=
\acos(1/3)$ and $\phi'=0$. That is, the best result is obtained by using a
$50/50$ beam splitter followed by a beam splitter with a reflectivity of $1/9$.

Thus we see that, if the inputs to an interferometer are in pure superposition
states, it is possible to obtain a perfect single-photon output for three modes.
In contrast, if the inputs to the interferometer are incoherent superpositions
of Fock states, it is impossible to obtain an improvement in the
single-photon probability for three modes \cite{BSSK}. These results
imply that it is the incoherence in the inputs that prevents an improvement in
the single photon probability. It would be interesting to determine the degree
of decoherence that is sufficient to prevent an improvement in the single-photon
efficiency. However, this is a difficult problem, which we leave to further
investigation.

\section{Multimode incoherent inputs}
\label{Sec:general}
Although it is possible to obtain perfect single-photon states from pure
superposition states, this method can not be applied to current experiments, as
single-photon sources do not produce pure superposition states. Real
single-photon sources produce an incoherent combination of Fock states;
therefore we consider input states of this form for the remainder of this paper.
In the multimode case we start with a supply of $N$ mixed states of the form 
\eqref{eq:initialstate}. For additional generality we allow the different inputs 
to have different probabilities for a single photon, $p_i$, and we denote the
maximum of these probabilities by $p_{\rm max}$. The initial input state may be
described by
\begin{align}
\label{eq:input}
\hat{\rho}_{\text{in}}^{(N)} & = \bigotimes_{i=1}^N \big[ (1-p_i) \ket 0 \bra 0
+ p_i \ket 1 \bra 1 \big] \nn \\    & = \sum_{\bm{s}} P_{\bm{s}} \left(
\prod_i (\hat a_i\dg)^{s_i} \ket{0}\bra{0} \prod_i (\hat a_i)^{s_i}\right),
\end{align}
where $P_{\bm{s}}=\prod_i p_i^{s_i}(1-p_i)^{1-s_i}$, and the vector
$\bm{s}$ $\!=$ $\!(s_1, \cdots , s_N)^T$, ($s_i=0,1$), gives the photon numbers
in the inputs. The quantity $P_{\bm{s}}$ is the probability of obtaining this
combination of input photon numbers.

\begin{figure}[t]
\centerline{\includegraphics[width=4.4cm]{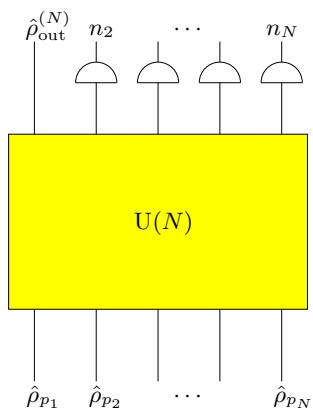}}
\begin{picture}(0,0)
\put(-55.5,3){$\hat{\rho}_{p_1}$}
\put(-32.1,3){$\hat{\rho}_{p_2}$}
\put(0,3){$\cdots$}
\put(37.8,3){$\hat{\rho}_{p_N}$}
\put(-56,142){$\hat{\rho}^{(N)}_{\text{out}}$}
\put(-32,142){$n_2$}
\put(0,142){$\cdots$}
\put(35,142){$n_N$}
\put(-15,68.5){U$(N)$}
\end{picture}
\caption{\label{fig:scheme} Schematic setup of the network. We assume $N$
incoming modes prepared in the state \eqref{eq:input} with different $p_i$.
The photon number is measured in output modes 2 to $N$, and we wish to improve
the probability for a single photon in mode 1.}
\end{figure}

This input is then passed through a passive interferometer which consists of
beam splitters, mirrors, and phase shifters. Each of these elements preserves
total photon number from input to output under ideal conditions. No energy is
required to operate these optical elements, hence the term passive (also known
as linear optical elements). More generally, polarisation transforming elements
can be included, but here we are concerned only with a scalar field treatment;
in fact polarisation effects could be included by doubling the number of
channels and treating the two polarisations in a mode as two separate channels.

Classically the field amplitude of channel $i$ would be represented by the
complex number $a_i$. The set of all field amplitudes for the $N$-channel
interferometer is given by the vector $\bm{a}=(a_1, \cdots ,a_N)^\text{T}$. The
passive interferometer transforms the input amplitudes to the output amplitudes
via the matrix transformation $\bm{a}\mapsto\bm{\Lambda}\dg \bm{a}$ with
$\Lambda\in\text{U}(N)$, where U$(N)$ is the set of all $N\times N$ unitary
matrices. Quantisation of the field is obtained by the replacement of $\bm{a}$
by the vector annihilation operator $\hat{\bm{a}}$, and the interferometer
transforms the operators according to
$\hat{\bm{a}}^\dagger\mapsto\bm{\Lambda}^{\rm T}\hat{\bm{a}}^\dagger$
\cite{VogelWelsch}. This transformation of the operators yields
\begin{equation}
\hat\rho_{\rm trans}^{(N)} = \sum_{\bm{s}} P_{\bm{s}} \left[ \prod_i \left(
\sum_k \Lambda_{ki} \hat a_k\dg\right)^{s_i} \ket{0}\bra{0} \prod_i \left(
\sum_k \Lambda_{ki}^* \hat a_k\right)^{s_i} \right].
\end{equation}

In the completely general case, we could perform photodetections on $N-N_1$
of the modes, and use the remaining $N_1$ modes as single-photon sources if
the desired combination of detection results is obtained. However, this
generality is not needed here because we are concerned with the maximum
improvement in the photon statistics in a single mode. In order to fix notation,
we denote as mode 1 that mode for which we want to improve the statistics,
and label the other modes where photodetections have not been
performed as modes 2 to $N_1$. The reduced density matrix in mode 1 is
then identical to what would be obtained if photodetections were
performed on modes 2 to $N_1$ (as well as $N_1+1$ to $N$),
and the results of these photodetections discarded. Therefore, the probability
for a single photon will be a weighted average of the single-photon
probabilities for the different combinations of detections in modes 2
to $N_1$. Hence the maximum single-photon probability in mode 1 will
be obtained for some combination of detections in modes 2 to $N_1$.
For this reason we consider the state in mode 1 conditioned on photodetections
in the other $N-1$ modes. As our aim is to determine the best results possible
using linear optics and photodetection, we also assume that the photodetectors
perform perfect photon counting measurements (imperfect detection is
discussed later).

Before determining the conditional output state, we introduce some additional
notation. The total number of photons detected is $D$, and the maximum possible
number of photons input to the interferometer is $M$. As some of the $p_i$ may
be equal to zero, $M$ may be less than $N$; $M$ is equal to the number of
nonzero values of $p_i$. For $j>1$, $n_j$ is the number of photons detected in
mode $j$, and $n_1$ is the photon number in mode 1 (the output mode). We use the
notation $\Sigma_n = \sum_i n_i$ (so $\Sigma_n=D+n_1$) and $\Sigma_s = \sum_i
s_i$. In addition, we define the set $\Phi_{\bm{s}}=\{ i | s_i = 1 \}$, and let
${\rm Y}_{\bm{s}}$ be the set that consists of all vectors comprised of the
elements of $\Phi_{\bm{s}}$.

The conditional state in mode 1 after photodetection in modes 2 to $N$ is
\begin{equation}
\hat\rho_{\rm out}^{(N)} = \sum_{n_1=0}^N c_{n_1} \ket{n_1}\bra{n_1}.
\end{equation}
Each coefficient $c_{n_1}$ is given by
\begin{equation} \label{expect}
c_{n_1} = K\bra{\bm{n}}\hat\rho_{\rm trans}^{(N)}\ket{\bm{n}},
\end{equation}
where $\ket{\bm{n}}$ is a tensor product of number states in each of the output
modes and the normalisation constant $K$ is equal to
\begin{equation}
K=\left[\sum_{n_1=0}^N \bra{\bm{n}}\hat\rho_{\rm trans}^{(N)}
\ket{\bm{n}}\right]^{-1}.
\end{equation}
Evaluating $c_{n_1}$ gives
\begin{equation} \label{general}
c_{n_1} = \frac {K'}{n_1!} \sum_{\bm{s};\Sigma_{\bm{s}}=\Sigma_{\bm{n}}}
P_{\bm{s}}\left| S_{\bm{s},\bm{n}}\right|^2,
\end{equation}
where $K'=K/\prod_{j=2}^N n_j!$, and
\begin{equation}
S_{\bm{s},\bm{n}} = \sum_{\bm{\sigma}\in {\rm Y}_{\bm{s}}} \left(
\Lambda_{1,\sigma_1} \cdots \Lambda_{1,\sigma_{n_1}} \right)
\cdots \left( \Lambda_{N,\sigma_{\Sigma_{\bm{s}}-n_N+1}} \cdots 
\Lambda_{N,\sigma_{\Sigma_{\bm{s}}}} \right).
\end{equation}
This quantity may alternatively be expressed using permanents as
\begin{equation}
S_{\bm{s},\bm{n}} = \text{per} (\bm{\Lambda}[\bm{n},\bm{s}]).
\end{equation}
Here the notation $\bm{\Lambda}[\bm{n},\bm{s}]$ is used to indicate that the
$i$'th column of $\bm{\Lambda}$ is repeated $s_i$ times, and the $j$'th row
is repeated $n_j$ times.

Two figures of merit for an arbitrary single mode field
$\sum_i q_i \ket{i}\bra{i}$ are
\begin{equation}
R = \frac{q_1}{q_0}, \qquad G = \frac{q_2/q_1}{q_1/q_0}.
\end{equation}
We use the subscript ``out'' to indicate the output field, and ``in'' to
indicate the input field. For the output field we simply have $q_i=c_i$. For the
input field, we have $G_{\rm in}=0$, as the two-photon component is assumed be
negligible. For simplicity we define $R_{\rm in}$ to be the maximum input
ratio $p_{\rm max}/(1-p_{\rm max})$.

The figure of merit $G$ characterises the two-photon contribution, and
is equal to $1/2$ for Poisson photon statistics.
If the multiphoton component in the output is zero, then comparing $R_{\rm in}$
and $R_{\rm out}$ immediately tells us if there is an improvement in the
probability for a single photon. Even if the multiphoton component is nonzero,
using $R_{\rm out}$ has the advantages: \\
1. The common constant $K'$ cancels, so it is possible to evaluate $R_{\rm out}$
analytically. \\
2. If $R_{\rm out}\le R_{\rm in}$, then it is clear that $c_1\le p_{\rm max}$.
Thus we can determine those cases where there is {\it no} improvement. \\
3. For $p_{\rm max}\ll 1$, $c_0\approx 1$ and $R_{\rm in}\approx p_{\rm max}$.
Therefore the improvement in $R$ is approximately the same as the improvement in
the single-photon probability over $p_{\rm max}$.

An alternative measure of the multiphoton contributions is given by how
sub-Poissonian the field is. That is, we may define the measure
\begin{equation}
\Pi = \frac{\langle \hat n^2 \rangle - \langle \hat n \rangle^2}{\langle \hat n
\rangle}.
\end{equation}
For a sub-Poissonian field, $\Pi<1$. States of the form \eqref{eq:initialstate}
are sub-Poissonian, with $\Pi = 1-p$. We take $\Pi_{\rm in}$ to be the minimum
value in the inputs, $1-p_{\rm max}$, and $\Pi_{\rm out}$ is simply the value
for the output mode. If an output $c_1$ greater than $p_{\rm max}$ is obtained,
while maintaining a multiphoton contribution that is zero or very small, then it
is clear that $\Pi_{\rm out}$ will be smaller than $\Pi_{\rm in}$. On the other
hand, if the output has multiphoton contributions similar to those for a Poisson
distribution, $\Pi_{\rm out}$ will be closer to 1.

\section{Limit on improvement}
\label{Sec:limit}
Ideally we wish to obtain an improvement in the figure of merit $R$, while
maintaining a value of $G$ that is zero, or at least small with respect to $1/2$
(the value for a Poisson distribution). This is a difficult task, so for
simplicity we begin by focusing on improving $R$. As was shown in
Ref.\ \cite{BSSK}, there is an upper limit on how far $R$ can be increased. Here
we show this result in more detail.

First we consider the expression for $c_0$:
\begin{equation}
\label{c0}
c_0 = K' \sum_{\bm{s};\Sigma_{\bm{s}}=D}
P_{\bm{s}}\left| S_{\bm{s},\bm{n}^{0}}\right|^2.
\end{equation}
Here $n_1^0=0$ and $n_j^0$ is the combination of detection results for $j>1$.
A simple way of re-expressing this summation is
\begin{equation}
\label{altno}
c_0 = \frac {K'}{N-D} \sum_{\bm{s}; \Sigma_{\bm{s}}=D+1}
\sum_{k;s_k=1} P_{\bm{s}^k} \left| S_{\bm{s}^k,\bm{n}^0} \right|^2,
\end{equation}
where $s^k_i=s_i$ except for $s^k_k=0$. That is, we consider combinations of
input photons $\bm{s}$ with one too many photons, then remove one of these
photons to obtain the correct number of input photons.
This expression for the sum gives each term $N-D$ times, so it is necessary
to divide by $N-D$ to obtain the correct result. Specifically, each alternative
$\bm{s}^k$ may be obtained from an $\bm{s}$ which is identical, except one of
the zeros of $\bm{s}^k$ is replaced with a one. As each $\bm{s}^k$ has $N-D$
zeros, there are $N-D$ possible alternative $\bm{s}$ that give the same
$\bm{s}^k$. 

If some of the inputs to the interferometer are simply vacuum states
(i.e.\ some of the $p_i$ are zero), it is possible to express the summation
in a more efficient way. First note that those terms in the sum in
Eq.\ \eqref{c0} with $P_{\bm{s}}=0$ do not contribute to the sum; therefore
we may restrict to terms with $P_{\bm{s}}\ne 0$.
\begin{equation}
c_0 = K' \sum_{\substack{\bm{s}; P_{\bm{s}}\ne 0 \\
\Sigma_{\bm{s}}=D }} P_{\bm{s}}\left| S_{\bm{s},\bm{n}^{0}}\right|^2.
\end{equation}
We may re-express this equation as
\begin{equation}
\label{c0alt}
c_0 = \frac {K'}{M-D} \sum_{\substack{\bm{s}; P_{\bm{s}}\ne 0 \\
\Sigma_{\bm{s}}=D+1 }} \sum_{k;s_k=1} P_{\bm{s}^k} \left| S_k \right|^2,
\end{equation}
where $S_k = S_{\bm{s}^k,\bm{n}^0}$. Here the dividing factor required is only
$M-D$.

Recall that the maximum total number of photons is $M$, so there are
$N-M$ inputs with $p_i=0$. Each alternative $\bm{s}^k$ still has $N-D$ zeros,
but some of these zeros will correspond to inputs with $p_i=0$. Because we are
restricting to terms with $P_{\bm{s}}\ne 0$, for all $i$ with $p_i=0$, $s_i=0$
and therefore $s_i^k$ must be equal to zero. Thus all $N-M$ of the
inputs with $p_i=0$ must correspond to zeros of $\bm{s}^k$, and so there
will only be $M-D$ zeros of $\bm{s}^k$ that correspond to nonzero $p_i$.
As in the previous case, each $\bm{s}^k$ may be obtained from an $\bm{s}$
which is identical, except one of the zeros of $\bm{s}^k$ is replaced with a
one. However, because we have restricted to terms with $P_{\bm{s}}\ne 0$,
the zero that is replaced with a one must be for an $i$ with nonzero $p_i$.
Hence there are only $M-D$ alternative $\bm{s}$ that lead to the same
$\bm{s}^k$, and the redundancy in this case is only $M-D$. That is why a
dividing factor of $M-D$ is required in this case.

Simplifying Eq.\ \eqref{c0alt}, we obtain
\begin{align}
c_0 &= \frac {K'}{M-D} \sum_{\substack{\bm{s}; P_{\bm{s}}\ne 0 \\
\Sigma_{\bm{s}}=D+1 }} P_{\bm{s}} \sum_{k;s_k=1} \frac{1-p_k}{p_k} \left| S_k
\right|^2.
\end{align}
Because the sum is limited to terms where $P_{\bm{s}} \ne 0$, $p_k$ is nonzero,
and therefore the ratio $(1-p_k)/p_k$ does not diverge. Using
$p_k\le p_{\rm max}$, we obtain the inequality
\begin{equation} \label{ineq0}
c_0 \ge \frac {K'/R_{\rm in}}{M-D} \sum_{\bm{s};
\Sigma_{\bm{s}}=D+1} P_{\bm{s}} \sum_{k;s_k=1} \left| S_k \right|^2.
\end{equation}
We now allow terms with $P_{\bm{s}}=0$, because they do not contribute to the
sum.

It is also possible to obtain an inequality for $c_1$. The probability $c_1$
is given by
\begin{equation}
\label{c1}
c_1 = K' \sum_{\bm{s};\Sigma_{\bm{s}}=D+1}
P_{\bm{s}}\left| S_{\bm{s},\bm{n}^{1}}\right|^2.
\end{equation}
In this case the notation $\bm{n}^1$ means $n_1^1=1$, and $n_j^1$ is the
combination of detection results for $j>1$. We may express
$S_{\bm{s},\bm{n}^{1}}$ as
\begin{align}
S_{\bm{s},\bm{n}^{1}} &= \sum_{\bm{\sigma}\in {\rm Y}_{\bm{s}}}
\Lambda_{1,\sigma_1}\left(\Lambda_{2,\sigma_2} \cdots
\Lambda_{2,\sigma_{n_2+1}} \right) \cdots \left(
\Lambda_{N,\sigma_{\Sigma_{\bm{s}}-n_N+1}} \cdots
\Lambda_{N,\sigma_{\Sigma_{\bm{s}}}} \right) \nn \\
&= \sum_{k;s_k=1} \Lambda_{1k} \sum_{\bm{\sigma}\in {\rm Y}_{\bm{s}^k}}
\left(\Lambda_{2,\sigma_1} \cdots \Lambda_{2,\sigma_{n_2}} \right)
\cdots \left( \Lambda_{N,\sigma_{\Sigma_{\bm{s}^k}-n_N+1}} \cdots 
\Lambda_{N,\sigma_{\Sigma_{\bm{s}^k}}} \right) \nn \\ &= \sum_{k;s_k=1}
\Lambda_{1k} S_{\bm{s}^k,\bm{n}^{0}} \nn \\ &= \sum_{k;s_k=1} \Lambda_{1k} S_k.
\end{align}
Therefore, we may re-express the equation for $c_1$ as
\begin{equation} \label{onephoton}
c_1 = K' \sum_{\bm{s};\Sigma_{\bm{s}}=D+1} P_{\bm{s}}
\left| \sum_{k;s_k=1} \Lambda_{1k} S_k \right|^2.
\end{equation}
This then gives the inequality
\begin{equation} \label{ineq1}
c_1 \le K' \sum_{\bm{s};\Sigma_{\bm{s}}=D+1 } P_{\bm{s}} \sum_{k;s_k=1}
\left| S_k \right|^2.
\end{equation}

Combining Eqs.\ \eqref{ineq0} and \eqref{ineq1}, we can see that
\begin{equation} \label{ineq1a}
R_{\rm out} = \frac{c_1}{c_0}\le R_{\rm in}(M-D) .
\end{equation}
This yields an upper limit on the ratio between the probabilities for one and
zero photons. This result allows one to draw three main conclusions: \\
1. As $M\le N$ and $D\ge 0$, the improvement in $R$ can never be greater than
$N$. There is no known scheme that saturates this upper bound, but there is a
scheme known that achieves an improvement of approximately $N/4$ \cite{BSSK}. \\
2. If the number of photons detected is one less than the maximum input number,
then $M-D=1$, and there can not be an improvement\footnote{This no-go theorem
has also been shown by E.\ Knill for the case that $M=N$ (personal
communication).}. This case is important because it is the most straightforward
way of eliminating the possibility of two or more photons in the output mode. We
have not proven that it is impossible to obtain an improvement while eliminating
the multiphoton component, but if such a scheme is possible it can not eliminate
the multiphoton component by detecting one less than the maximum input number of
photons. \\
3. It is impossible to obtain a single photon with unit probability if
$p_{\rm max}<1$. If $c_1=1$ were obtained, then $R_{\rm out}$ would be infinite;
from Eq.\ \eqref{ineq1a}, this is clearly not possible unless $R_{\rm in}$ is
infinite (which would correspond to $p_{\rm max}=1$).

\section{Method for improvement}
\label{Sec:method}
We have previously shown that it is possible to obtain an improvement in the
probability for a single photon \cite{BSSK}. Here we review this method, giving
more motivation for this scheme and propose a simple realisation using a line of
beam splitters. In order to obtain a value for the ratio $R_{\rm out}$ that is
close to the upper limit, we require the two inequalities \eqref{ineq0} and
\eqref{ineq1} to be as close to equality as possible. We may achieve equality in
the first case \eqref{ineq0} by taking all nonzero $p_i$ equal to $p_{\rm max}$.

To obtain equality in Eq.\ \eqref{ineq1}, we would require $s_k=1$ whenever
$\Lambda_{1k}$ is nonzero, and $S_k$ to be proportional to $\Lambda_{1k}^*$
for those values of $k$ where $s_k=1$. These conditions would need to be
satisfied for all $\bm{s}$ that give nonzero $P_{\bm{s}}$. The first condition
is a problem, because it can not be satisfied unless $D=M-1$. To see this,
note that this condition is equivalent to requiring that $s_k=0$ implies
$\Lambda_{1k}=0$, for all $\bm{s}$ that give $P_{\bm{s}}>0$. If
$P_{\bm{s}}>0$, then $s_k=0$ for all $k$ such that $p_k=0$. Therefore,
$p_k=0$ implies $\Lambda_{1k}=0$. Now if $D+1<M$, then it must be the case
that $s_k=0$ for some $k$ such that $p_k>0$. In addition, for arbitrary $k$,
there will be an $\bm{s}$ with $P_{\bm{s}}>0$ such that $s_k=0$ and $p_k>0$.
Therefore, for the first condition to be satisfied, it would be necessary for
$\Lambda_{1k}$ to be equal to zero for all $k$. This is clearly not possible,
because $\bm{\Lambda}$ is a unitary matrix.

On the other hand, if $D+1=M$, then the only $\bm{s}$ giving $P_{\bm{s}}>0$
is that with $s_k=1$ for $p_k>0$, and $s_k=0$ for $p_k=0$. Therefore, it is
possible for the first condition to be satisfied, by choosing a $\bm{\Lambda}$
with $\Lambda_{1k}=0$ for all $k$ such that $p_k=0$. However, the case with
$D=M-1$ is unimportant, because it is not possible to obtain an improvement in
$R$. Hence we see that, in any case where it is possible to obtain
an improvement, it is not possible to obtain equality in Eq.\ \eqref{ineq1a}.

On the other hand, we can determine a scheme that gives
$S_k\propto\Lambda_{1k}^*$. This condition can be satisfied using the
interferometer with matrix elements
\begin{align} \label{interfer}
\Lambda_{11} =-\epsilon, & \qquad \Lambda_{21} = \sqrt{1-\epsilon^2}, \nn \\
\Lambda_{1i} =\sqrt{(1-\epsilon^2)/(N-1)}, & \qquad
\Lambda_{2i}=\epsilon/\sqrt{N-1}, 
\end{align}
for $i>1$ (the values of $\Lambda_{ij}$ for $i>2$ do not enter into the
analysis). Here $\epsilon$ is a small number, and we ignore terms of order
$\epsilon$ or higher. Now let $p_i=p_{\rm max}$, and consider the measurement
record where zero photons are detected in modes 3 to $N$, so the number of
photons detected in mode 2 is $D$.

\begin{figure}[t]
\centerline{\includegraphics[width=5.5cm]{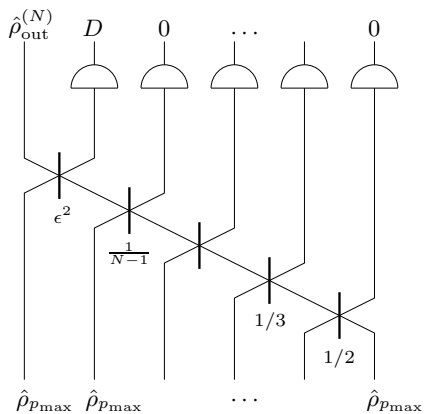}}
\begin{picture}(0,0)
\put(-81.1,1){$\hat{\rho}_{p_{\rm max}}$}
\put(-54.8,1){$\hat{\rho}_{p_{\rm max}}$}
\put(0,1){$\cdots$}
\put(51.4,1){$\hat{\rho}_{p_{\rm max}}$}
\put(-84,142){$\hat{\rho}^{(N)}_{\text{out}}$}
\put(-55.9,141){$D$}
\put(-27.3,141){$0$}
\put(0,140){$\cdots$}
\put(52.2,141){$0$}
\put(-67,70){\scriptsize $\epsilon^2$}
\put(-47,57){\scriptsize $\frac 1{N-1}$}
\put(9,31){\scriptsize $1/3$}
\put(35,17){\scriptsize $1/2$}
\end{picture}
\caption{\label{fig:imp} A realisation of the interferometer specified in
Eq.\ \eqref{interfer} using beam splitters. The reflectivities of the beam
splitters are specified in the labels on the beam splitters. This example is for
$N=6$.}
\end{figure}

This scheme is the same as in Ref.\ \cite{BSSK}, except input modes 1 and 2
have been swapped. Expressing the scheme in this form allows us to determine a
simple realisation using beam splitters. This scheme may be performed using the
chain of beam splitters shown in Fig.\ \ref{fig:imp}. The first $N-2$ beam
splitters (those on the right) result in an output beam with equal contributions
from $N-1$ of the inputs. This equal combination is achieved by decreasing the
reflectivities from $1/2$ for the first beam splitter to $1/(N-1)$ for beam
splitter $N-2$. The last beam splitter has the low reflectivity $\epsilon^2$.
With appropriate phase shifts, these beam splitters give the overall
interferometer described by $\bm{\Lambda}$ in Eq.\ \eqref{interfer}.

To determine $c_{n_1}$, note first that
$\Lambda_{21}\gg \Lambda_{2i}$ for $i>1$, so we may ignore those terms in the
sum for $S_{\bm{s},\bm{n}}$ where $\Lambda_{21}$ does not appear. Each term has
magnitude $\Lambda_{12}^{n_1}\Lambda_{21}\Lambda_{22}^{D-1}$ \footnote{We
use $\Lambda_{12}$ and $\Lambda_{22}$ to indicate the values of
$\Lambda_{1i}$ and $\Lambda_{2i}$ for $i>1$.}, and there are $D(D+n_1-1)!$
such terms. Therefore, provided $s_1=1$,
\begin{equation}
\label{Ssn}
S_{\bm{s},\bm{n}} \approx D(D+n_1-1)!\Lambda_{12}^{n_1}\Lambda_{21}
\Lambda_{22}^{D-1}. 
\end{equation}
If $s_1=0$, then $S_{\bm{s},\bm{n}}$ is of order $\epsilon$.

Recall that $S_k=S_{\bm{s}^k,\bm{n}^0}$, where $\bm{s}^k$ is equal to $\bm{s}$,
except for $s_k^k=0$, and $n_1^0=0$. (We do not consider $k$ where $s_k=0$.)
The result for $k>1$ may be obtained by replacing $n_1$ with 0 and $D$ with
$\Sigma_{\bm{s}}-1$ in Eq.\ \eqref{Ssn}, giving
\begin{equation}
S_k \approx (\Sigma_{\bm{s}}-1)(\Sigma_{\bm{s}}-2)!\Lambda_{21}
\Lambda_{22}^{\Sigma_{\bm{s}}-2}.
\end{equation}
For $k=1$, we simply obtain $S_1$ of order $\epsilon$. Similarly,
$\Lambda_{1k}$ is constant for $k>1$, and of order $\epsilon$ for $k=1$.
Thus this scheme gives $S_k\propto\Lambda_{1k}^*$, as claimed above.

In order to determine $c_{n_1}$, note that there are $\binom{N - 1}{D+n_1-1}$
different combinations of inputs such that $\Sigma_{\bm{s}}=D+n_1$ and $s_1=1$.
Combining this expression with Eq.\ \eqref{Ssn}, we have
\begin{align} \label{total}
c_{n_1} &\approx \frac{K'}{n_1!}p_{\rm max}^{D+n_1}(1-p_{\rm max})^{N-D-n_1}
 \frac{(N-1)!D^2(D+n_1-1)!}{(N-D-n_1)!}\Lambda_{12}^{2n_1}
\Lambda_{21}^2\Lambda_{22}^{2D-2} \nn \\ &= K'' \left(\frac{R_{\rm in}}
{N-1}\right)^{n_1} \frac{(D+n_1-1)!}{n_1!(N-D-n_1)!}.
\end{align}
We have combined those factors that do not depend on $n_1$ into a new constant
$K''$, and used $\Lambda_{12}\approx 1/\sqrt{N-1}$.
Using Eq.\ \eqref{total} gives
\begin{equation}
R_{\rm out} \approx R_{\rm in} \frac{D(N-D)}{N-1}.
\end{equation}
The maximum improvement in $R$ is obtained for $D=\lceil N/2 \rceil$,
where $R_{\rm out}\approx R_{\rm in}\lfloor N^2/4 \rfloor/(N-1)$. The
multiplicative factor $\lfloor N^2/4 \rfloor/(N-1)$ is larger than 1 for all
$N\ge 4$. Thus we find that, provided there are at least 4 modes, we may obtain
an improvement in $R_{\rm out}$. For $p_{\rm max}\ll 1$, $c_1 \approx
p_{\rm max}\lfloor N^2/4\rfloor/(N-1)$. For large $N$, the probability of a
single photon increases approximately as $N/4$, but does not achieve the upper
bound of $N$.

Although we find an improvement in the measure $R$, the two-photon contribution
is not negligible. Using the measure $G$, we find
\begin{equation}
G_{\rm out} = \frac{c_2/c_1}{c_1/c_0} \approx \frac{(D+1)(N-D-1)}{2D(N-D)} .
\end{equation}
For $D=\lceil N/2 \rceil$, this measure is close to $1/2$, so the two-photon
component is similar to that for a Poisson distribution. By taking $D=N-2$, it
is possible to obtain an improvement in $R$ of about a factor of two,
with a value of $G_{\rm out}$ about half that for a Poisson distribution.
However, this two-photon contribution is still much greater than for good
single-photon sources \cite{twophot}.

The multiphoton contributions are especially important for larger $p_{\rm max}$.
Although the improvement in $R$ is independent of $p_{\rm max}$, the
multiphoton component means that improvements in $c_1$ are obtained only for
values of $p_{\rm max}$ below $1/2$. That is, this method can only be used to
obtain improvements in the probability of a single photon up to $1/2$, but not
to make the probability of a single photon arbitrarily close to 1.

This scheme also performs poorly when evaluated via the measure $\Pi$. It is
more difficult to evaluate this scheme using this measure, and we do not have a
simple solution for $\Pi$. However, numerically we find that $\Pi_{\rm out}$ is
close to 1 for $D=\lceil N/2 \rceil$, again indicating that the output state is
close to Poissonian. For $D=N-2$, $\Pi_{\rm out}$ is closer to $\Pi_{\rm in}$
($=1-p_{\rm max}$), but for no $D$ is a value of $\Pi_{\rm out}$ less than
$\Pi_{\rm in}$ obtained.

Nevertheless, this scheme does give an improvement in $R$. It can be expected
that this improvement is close to the maximum possible, because this scheme
satisfies $S_k\propto\Lambda_{1k}^*$. However, note that it satisfies the other
condition for optimality, i.e.\ that $\Lambda_{1k}\ne 0$ implies $s_k=1$, fairly
poorly. As shown above, this condition can not be satisfied completely (unless
there is no improvement in $R$), and there does not appear to be any method of
satisfying it better than the method we have described above. Extensive
numerical searches have failed to find any scheme that gives a better
improvement in $R$ than the above scheme, strongly indicating that it is optimal
for increasing $R$.

\section{Experimental limitations}
\label{Sec:exp}
In practice, there will be a number of limitations to using this method for
improving the probability of a single photon. The main ones are: \\
1. Real photodetectors do not give perfect photon counting measurements. Most
photodetectors can only distinguish between the vacuum state and a state with
one or more photons. \\
2. Real photodetectors have limited efficiency and dark counts. \\
3. The desired combination of detection results will occur with low
probability. \\
4. Real sources will have a finite multiphoton component. \\

For $N-2$ of the detectors, point 1 will not be a problem. The reason for this
is that we are conditioning on detection of zero photons at these detectors. It
is only the detector on mode 2 that is required to perform a photon counting
measurement. Even for this detector, it is not necessary to determine the exact
photon number. That is because the probability of a single photon will be
increased for any number of photons detected from 2 to $N-2$. Therefore, if the
detector can register that the photon number is in this range, rather than the
exact photon number, it will be sufficient to produce an improvement in the
single-photon probability.

Alternatively, an improvement can also be achieved using detection that simply
verifies that there is more than one photon. For example, the Visible Light
Photon Counter \cite{VLPC} can do this task with high efficiency. Even though
the possibilities of $N-1$ or $N$ photons have not been eliminated, they have
lower probability, and will not contribute significantly to the photon
probabilities.

Limited efficiency is not a severe problem for small $p_{\rm max}$, because the
probability for the vacuum is relatively large. Dark counts will not be a
problem for the first $N-2$ detectors, because we are conditioning on vacuum
detection at these detectors. Dark counts will merely slightly reduce the
probability of obtaining the desired combination of detection results. On the
other hand, dark counts will be a problem for detector 2, as we are conditioning
on detection of more than one photon at this detector.

Point 3 will always be a problem, because the above scheme is only effective for
small $\epsilon$. The probability for this combination of detection results
becomes very small in the limit of small $\epsilon$. If larger values of
$\epsilon$ are used, then the final probability for a single photon becomes
smaller. Thus there is a trade-off involved. Point 4 is more complicated, and it
is not clear how important this problem is without performing direct
calculations.

\begin{figure}[t]
\centerline{\includegraphics[width=8cm]{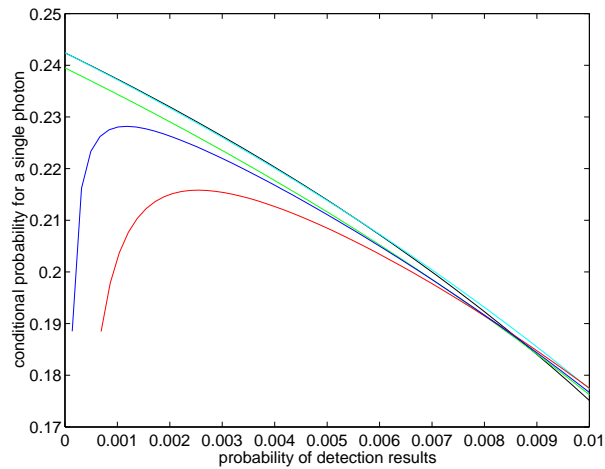}}
\caption{\label{fig:exp} The final probability for a single photon versus the
probability for obtaining the appropriate detection results for a four mode
interferometer and $p_{\rm max}=0.2$. The black line is the ideal case, the
light blue line is that for the case where detection results of $D=3$ and $D=4$
are also allowed, the green line is that if, in addition, the photodetectors
have 90\% efficiency. The dark blue line is that including all of these
experimental limitations, plus 0.1\% chance of dark counts at detector 2, and
the red line is that taking into account these experimental limitations, as well
as allowing 0.1\% probability for two photons in the inputs (without dark
counts).}
\end{figure}

To estimate the relative importance of each of the above problems, we have
calculated the final conditional probability for a single photon successively
taking each of the above issues into account. In Fig.\ \ref{fig:exp}, we have
plotted the conditional probability for a single photon versus the probability
for obtaining the desired combination of detection results. These curves are
parametrised by $\epsilon$; that is, both probabilities were calculated for a
range of values of $\epsilon$. In general, as $\epsilon$ is decreased, the
probability for obtaining the desired detection results decreases, and the final
conditional probability for a single photon increases. The particular example we
have shown is of a 4 mode interferometer where $p_{\rm max}=0.2$ for the inputs.

To perform these calculations, the density matrix was left unnormalised. The
trace of the density matrix at the end of the calculation then gives the
probability for that combination of detection results. For those cases where
two, three and four photons were not distinguished, the density matrices for
these three cases were simply added. To take account of finite efficiency
detectors and dark counts, the density matrices for the other detection results
were multiplied by constant factors, and added to the density matrix for the
desired detection result. It was assumed that the inefficient detectors register
single-photon, two-photon and three-photon states as vacuum with probabilities
of 10\%, 1\% and 0.1\%, respectively. For the detector on mode 2 it was assumed
that a single-photon state is registered as two or more photons with 0.1\%
probability, and the vacuum state is registered as two or more photons with
0.0001\% probability (due to the lower probability of simultaneous dark counts).
The appropriate equations to use for the case with multiphoton inputs are
derived in Sec.\ \ref{Sec:multi}.

For perfect sources and detectors, the final probability for a single photon is
above the initial probability of 20\% when the probability for obtaining the
detection results is below about 0.7\%. Thus, in order to obtain the desired
detection results, the experiment needs to be repeated roughly 200 times, which
is not unreasonable. If we consider a final detector that can not distinguish
between two, three or four photons, the results are almost identical, so this
problem is relatively trivial.

Even photodetectors with finite efficiency do not greatly affect the results. If
the first two photodetectors have 90\% efficiency, the final probability for a
single photon is reduced by about 0.3\%. The greatest problems are dark counts
at detector 2, and multiphoton components at the inputs. If a dark count rate of
0.1\% is allowed, then the maximum single-photon probability is reduced below
0.23\%. For small values of $\epsilon$ the single-photon probability drops
dramatically, rather than approaching the maximum value. The results are similar
if a two-photon probability of 0.1\% is allowed in the inputs (while the vacuum
probability is decreased by 0.1\%). The single-photon probability again drops
for small values of $\epsilon$, and the maximum single-photon probability is
less than 0.22\%. For two-photon probabilities of 0.4\% or more it is not
possible to obtain any increase in the single-photon probability . Thus we see
that the main problems with experimental realisations will be two-photon
components in the input and dark counts at detector 2.

\section{No-go theorems}
\label{Sec:nogo}
In this section, we prove a number of no-go theorems for post-processing
via linear optics and photodetection. Note that one limitation of the scheme
given in Sec.\ \ref{Sec:method} is that it only gives improvements in $R$ for
four or modes. In fact, it is impossible to obtain improvements for fewer than
four modes \cite{BSSK}. This result may be shown in the following way. First
consider the case $D=0$. Then there is only one term in the sum for $c_0$, and
$c_0 = K' P_{\bm{0}}$. The expression for $c_1$ becomes
\begin{align}
c_1 & = K' \sum_{k=1}^N \frac {p_k}{1-p_k} P_{\bm{0}} \left| \Lambda_{1k}
\right| ^2  \nn \\ & 
\le K' R_{\rm in} \sum_{k=1}^N
P_{\bm{0}} \left| \Lambda_{1k} \right| ^2 \nn \\
& = K' R_{\rm in} P_{\bm{0}} \nn \\ &
= c_0 R_{\rm in}.
\end{align}
Thus we have shown that $R_{\rm out} \le R_{\rm in}$, so $c_1 \le p_{\rm max}$.
Hence there can be no improvement in the photon statistics if zero photons are
detected.

\begin{figure}[t]
\centerline{\includegraphics[width=4.4cm]{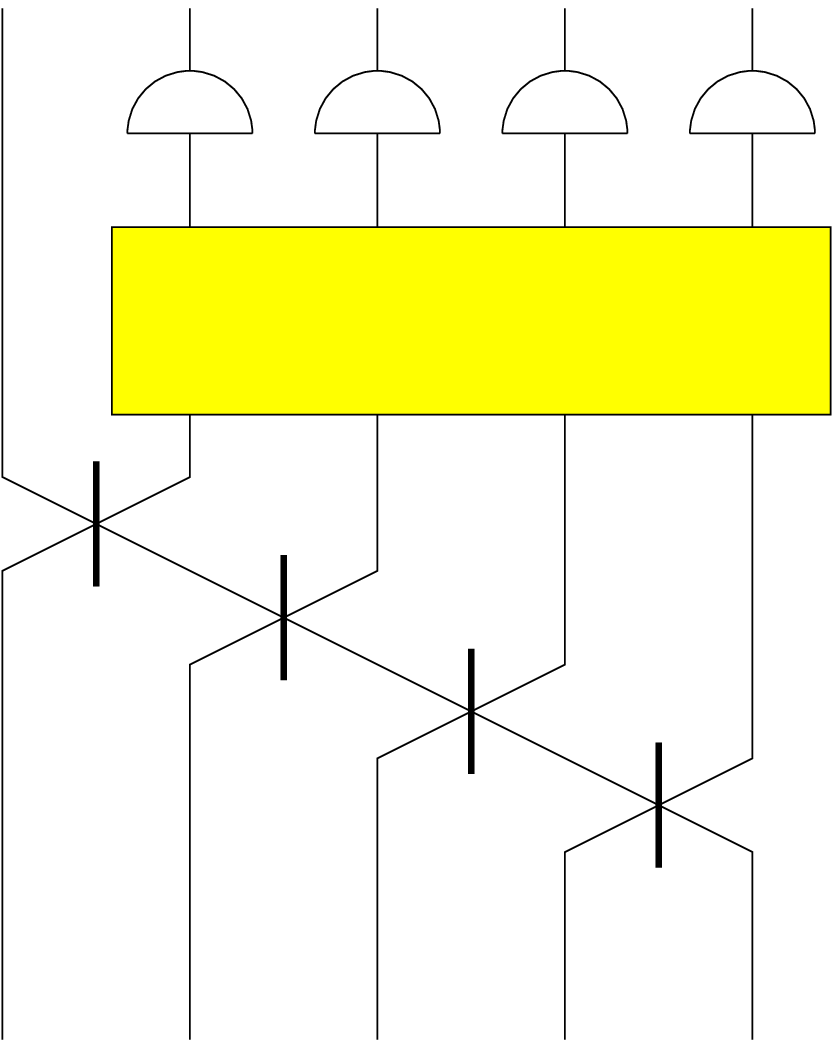}}
\begin{picture}(0,0)
\put(-65.5,3){$\hat{\rho}_{p_1}$}
\put(-39.8,3){$\hat{\rho}_{p_2}$}
\put(0,3){$\cdots$}
\put(36.8,3){$\hat{\rho}_{p_N}$}
\put(-67,156){$\hat{\rho}^{(N)}_{\text{out}}$}
\put(-39,155){$0$}
\put(0,155){$\cdots$}
\put(37.5,155){$0$}
\put(-15,108){U$(N-1)$}
\end{picture}
\caption{\label{fig:line} A U($N$) interferometer can be represented
by a U($N-1$) interferometer preceded by $N-1$ beam splitters. This
figure shows the example for $N=5$.}
\end{figure}

This result can also be shown in a more intuitive way as follows.
First note that an arbitrary U($N$) interferometer can be obtained using a
line of $N-1$ beam splitters followed by a U($N-1$) interferometer
(Fig.~\ref{fig:line}).
This is an immediate consequence of the algorithmic construction of
arbitrary U($N$) interferometers from beam splitters \cite{Reck94}.
If the $N-1$ modes upon which the U($N-1$) interferometer acts are
those that are measured, then we may omit the U($N-1$) interferometer
entirely (because detecting zero photons at the output of this
interferometer is identical to detecting zero photons at the
input). Thus this case may be reduced to the case of a line of beam
splitters where zero photons are detected at each stage. 

The case of a line of beam splitters with vacuum detection may be deduced from
the case for a single beam splitter. As was shown above, with a single beam
splitter there is no improvement in the ratio between the probabilities for
detecting one and zero photons. It is easily seen that the same result holds if
there are nonzero probabilities for photon numbers larger than 1 in the inputs
(corresponding to photon numbers larger than one in the output).

Thus, if we have a line of beam splitters, the ratio between the
probabilities for one and zero photons in the output can not be increased
above the maximum of that for the inputs. This result implies that the
probability of one photon in the output can never exceed $p_{\rm max}$.
This result holds for a line of beam splitters, and therefore for an
arbitrary U($N$) interferometer.

We can also obtain a similar result for the case $D=1$, provided all the input
$p_i$ are equal. If the single photon is detected in mode $m$, then 
\begin{equation}
c_0 = K' \sum_{k} \frac{p_{\rm max}}{1-p_{\rm max}} P_{\bm{0}}
\left| \Lambda_{mk} \right|^2 = K' R_{\rm in} P_{\bm{0}}.
\end{equation}
The value of $c_1$ is given by
\begin{align}
c_1 &= \frac 12 K' \sum_{k}\sum_{l;l\ne k}R_{\rm in}^2 P_{\bm{0}}
\left| \Lambda_{1l}\Lambda_{mk} +\Lambda_{1k}\Lambda_{ml} \right|^2 \nn \\
& \le \frac 12 K' R_{\rm in}^2 P_{\bm{0}} \sum_{k,l}
\left| \Lambda_{1l}\Lambda_{mk} + \Lambda_{1k}\Lambda_{ml} \right|^2 \nn \\
& = \frac 12 K' R_{\rm in}^2 P_{\bm{0}} \sum_{k,l} \left[
|\Lambda_{1l}|^2 |\Lambda_{mk}|^2 + |\Lambda_{1k}|^2 |\Lambda_{ml}|^2 + 
\Lambda_{1l}\Lambda_{mk} \Lambda_{1k}^*\Lambda_{ml}^* +
\Lambda_{1l}^*\Lambda_{mk}^* \Lambda_{1k}\Lambda_{ml}\right] \nn \\
& = \frac 12 K' R_{\rm in}^2 P_{\bm{0}} \left[ \sum_l 
|\Lambda_{1l}|^2 \sum_k |\Lambda_{mk}|^2 + \sum_k |\Lambda_{1k}|^2 \sum_l
|\Lambda_{ml}|^2 + \sum_l \Lambda_{1l}\Lambda_{ml}^* \sum_k \Lambda_{mk}
\Lambda_{1k}^* + \sum_l \Lambda_{1l}^*\Lambda_{ml}  \sum_k \Lambda_{mk}^*
\Lambda_{1k}\right] \nn \\ & = K' R_{\rm in}^2 P_{\bm{0}}.
\end{align}
In the last line we have used the fact that $\Lambda_{1k}$ and $\Lambda_{mk}$
are orthonormal. Thus we again find $R_{\rm out}\le R_{\rm in}$, so
$c_1 \le p_{\rm max}$.

These results can be used for an alternative proof that no improvement is
possible for the case of a single beam splitter. We have shown that detecting
zero photons does not give an improvement, and if one photon is detected, then
we must have $M-D=1$ or 0, so there again can be no improvement.

We can also eliminate the case of a three-mode interferometer, though the
reasoning is not as straightforward. First note that an input with probability
$p_i$ of a photon can be obtained by randomly selecting between a source with
efficiency $p_{\rm max}$ and the vacuum. That is, with probability
$q=p_i/p_{\rm max}$ we use the source with efficiency $p_{\rm max}$, and with
probability $1-q$ we use the vacuum state. If we discard the information about
which source was used, this is obviously equivalent to a source with efficiency
$p_i$. Hence the value of $c_1$ for the source with efficiency $p_i$ is the
weighted average of the values of $c_1$ for the cases where the efficiency is
$p_{\rm max}$ and zero. Thus the maximum $c_1$ must be obtained with all of the
nonzero $p_i$ equal to $p_{\rm max}$.

Therefore, in considering the three-mode interferometer, we can let all
the nonzero $p_i$ be equal to $p_{\rm max}$. If all the $p_i$ are
nonzero, then we can use the result showing that there is no improvement
with one photon detected and all $p_i$ equal. The other alternatives for
detection have $D=0$ or $M-D\le 1$, so there can be no improvements in
these cases either. If one or more of the $p_i$ are zero, then the only
detection alternatives are with $D=0$ or $M-D\le 1$. Thus we have shown
that there can never be an increase in the probability of a single photon
if less than four modes are used.

\section{Unsolved problems}
\label{Sec:conn}
Although it seems that we were able to answer most of the relevant
questions concerning the possibility of improving the efficiency of
single-photon sources, there are, in fact, still a number of open
questions we would now like to address. The two main unsolved
problems for this post-processing are: \\
1. Is it possible to increase the probability for a single photon,
regardless of the value of $p_{\rm max}$? \\
2. Can the single-photon probability be increased without adding a
multiphoton component? \\

At this time the indications are that the answer to both these questions is
no. We have performed numerical searches for interferometers that give
improvements for $p\ge 1/2$. These searches have been unsuccessful, indicating
that it is not possible to obtain an improvement for $p\ge 1/2$. We have not
been able to prove this assertion;
however, we can show that there are various implications if there is any
value of $p_{\rm max}$ such that it is impossible to obtain an improvement.

First note that it is sufficient to use $p_i=p_{\rm max}$ in the
input modes. As discussed above, for a given interferometer the maximum
improvement will always be obtained with all of the nonzero $p_i$
equal to $p_{\rm max}$. It is possible to obtain a vacuum state from
inputs with efficiency $p_{\rm max}$, simply by using a beam splitter
and conditioning on detection of two photons at one of the outputs.
Therefore, if there is an interferometer that achieves a certain result
using inputs with $p_i=0$ or $p_{\rm max}$, there will always be another
(expanded) interferometer that achieves the same result with
$p_i=p_{\rm max}$.

Using this simplification, the expression for $c_{n_1}$ simplifies to
$c_{n_1} = K''d_{n_1}R_{\rm in}^{n_1}/{n_1!}$, where
\begin{equation}
d_{n_1} = \sum_{\bm{s}; \Sigma_{\bm{s}}=\Sigma_{\bm{n}}}
\left| S_{\bm{s},\bm{n}} \right|^2.
\end{equation}
The values of $d_{n_1}$ are independent of $p_{\rm max}$, and depend
only on the interferometer and combination of detection results. There is
an improvement in the probability of a single photon if
\begin{equation}
\label{noimp}
d_1 > d_0 + \sum_{n_1=2}^N d_{n_1}R_{\rm in}^{n_1}/{n_1!} \, .
\end{equation}

Let $p_0$ be a value of $p_{\rm max}$ such that there is an improvement
in the probability of a single photon, and let the corresponding value
of $R_{\rm in}$ be $R_0$. Then there exists an interferometer and combination
of detection results such that
\begin{equation}
\label{p0imp}
d_1 > d_0 + \sum_{n_1=2}^N d_{n_1}R_0^{n_1}/{n_1!} \, .
\end{equation}
Since each of the $d_n$ are positive, and the right-hand side is increasing as
a function of $R_{\rm in}$, we find that \eqref{noimp} is satisfied for all
$0<p_{\rm max}\le p_0$. Thus we find that, for any value of $p_{\rm max}$ such
that there is an improvement, there is an improvement for all smaller values of
$p_{\rm max}$. In turn this result implies that, if there is no improvement for
$p_{\rm max}=p_0$, then there can be no improvement for larger values of
$p_{\rm max}$.

It is also possible to show that, if it were possible to obtain
an improvement with no multiphoton contribution, there would be no 
value of $p_{\rm max}$\footnote{Provided $p_{\rm max}$ is nonzero and less
than 1. These restrictions on $p_{\rm max}$ are implied in the following
text.} for which we can not obtain an improvement.
To show this, note that zero multiphoton contribution implies that
$d_{n_1}=0$ for $n_1\ge 2$. Therefore, if this improvement is possible
for $p_{\rm max}=p_0$, then Eq.\ \eqref{p0imp} becomes simply
\begin{equation}
d_1 > d_0.
\end{equation}
Similarly, the condition to obtain an improvement for any other value
of $p_{\rm max}$ is simply $d_1 > d_0$, which is automatically satisfied.
In addition, because $d_{n_1}=0$ for $n_1\ge 2$, $c_{n_1}=0$ for $n_1\ge 2$,
for any $p_{\rm max}$.

Thus, if it is possible to obtain an improvement for some value of
$p_{\rm max}$ while maintaining zero multiphoton contribution, then it
will be possible to obtain an improvement for all values of $p_{\rm max}$.
In addition, the relative improvement in $R$ is independent of $p_{\rm max}$.
To see this, note that
\begin{equation}
\frac{R_{\rm out}}{R_{\rm in}} = \frac{c_1/c_0}{R_{\rm in}} = \frac{d_1}{d_0}.
\end{equation}
A further implication is that it would be possible to obtain an output state
that is arbitrarily close to the pure single-photon state. As the output from
the interferometer has no multiphoton contribution, outputs from $N$ of these
interferometers may be used as the input to another, thus increasing $R$ by
a factor of $(d_1/d_0)^2$. Further iterations may be used to increase $R$
by a factor of $d_1/d_0$ to any arbitrary power, thus obtaining a final
probability for a single photon that is arbitrarily close to 1.

At this stage there is no known scheme that can give an improvement in the
probability for a single photon while maintaining zero multiphoton contribution.
As shown above, if this were possible for any value of $p_{\rm max}$, then it
would be possible for all values $p_{\rm max}$. As increasing the single-photon
probability {\it without} maintaining zero multiphoton component is a less
difficult problem, if it were possible to obtain an improvement while
maintaining zero multiphoton component, for all values $p_{\rm max}$ there would
be an enormous range of schemes that give improvements without the constraint on
the multiphoton component. Such a wide range of schemes would be relatively easy
to find numerically; the fact that numerical searches have failed to find any
scheme that gives an improvement in the single-photon probability for
$p_{\rm max}\ge 1/2$ therefore implies that it is extremely unlikely that there
is any scheme that gives an improvement while maintaining zero multiphoton
contribution. Nevertheless, these numerical results are not sufficient to rule
out this possibility.

\section{Multiphoton inputs}
\label{Sec:multi}
The majority of this study is based upon inputs from photon sources that have
zero multiphoton contribution. It is also possible to derive results for
inputs with nonzero probabilities for two or more photons, but this case is
more difficult. The simplest case is for a beam splitter with multiphoton
inputs. Let us denote the probability for $m$ photons in input mode $i$
by $p_{im}$. Then the input state may be written as
\begin{align}
\hat{\rho}_{\text{in}}^{(N)} &= \sum_{k,l} p_{1k} p_{2l} \ket{kl}\bra{kl} \nn \\
& = \sum_{k,l} \frac{p_{1k} p_{2l}}{k!l!} (a_1\dg)^k (a_2\dg)^l \ket{00}\bra{00}
(a_1)^k (a_2)^l .
\end{align}
The beam splitter transformation \eqref{splitter} gives
\begin{equation}
\hat\rho_{\rm trans}^{(N)} = \sum_{k,l} \frac{p_{1k} p_{2l}}{k!l!}
(\Lambda_{11}a_1\dg+\Lambda_{21}a_2\dg)^k (\Lambda_{12}a_1\dg+\Lambda_{22}
a_2\dg)^l \ket{00}\bra{00}(\Lambda_{11}^*a_1+\Lambda_{21}^*a_2)^k
(\Lambda_{12}^*a_1+\Lambda_{22}^*a_2)^l.
\end{equation}
Expanding in a series and conditioning upon detection of $D$ photons gives
\begin{align}
\hat\rho_{\rm out}^{(N)} &= K \sum_{k,l}p_{1k} p_{2l} k!l!D!(k+l-D)! \left| 
\sum_{m=\max(D-k,0)}^{\min(D,l)} \frac{\Lambda_{11}^{D-m}\Lambda_{12}^m
\Lambda_{21}^{k-D+m}\Lambda_{22}^{l-m}}{(D-m)!m!(k-D+m)!(l-m)!}\right|^2 \nn \\
&= K D! \frac{|\Lambda_{22}|^{2D}}{|\Lambda_{12}|^{2D}} \sum_{k,l}p_{1k} p_{2l}
k!l!(k+l-D)! |\Lambda_{12}|^{2k} |\Lambda_{22}|^{2l} \left|
\sum_{m=\max(D-k,0)}^{\min(D,l)} \frac{(-1)^m |\Lambda_{12}/\Lambda_{22}|^{2m}}
{(D-m)!m!(k-D+m)!(l-m)!}\right|^2 ,
\end{align}
where $K$ is a normalisation constant. If $K$ is omitted, the trace gives the
probability for this detection result. This is the expression used to calculate
the numerical results in Sec.\ \ref{Sec:exp}.

The transformation for a multimode interferometer is also straightforward to
determine. For this case the input state may be written as
\begin{align}
\hat{\rho}_{\text{in}}^{(N)} & = \bigotimes_{i=1}^N \big[ 
\sum_m p_{im} \ket{m}\bra{m} \big] \nn \\
& = \sum_{\bm{s}} \frac{P_{\bm{s}}}{\prod_i s_i!} \left( \prod_i
(\hat a_i\dg)^{s_i} \ket{0}\bra{0} \prod_i (\hat a_i)^{s_i} \right),
\end{align}
where $P_{\bm{s}}=\prod_i (p_{i,s_i})^{s_i}$. From this point on we use the
abbreviated notation $P'_{\bm{s}}=P_{\bm{s}}/\prod_i s_i!$. Note that $s_i$ may
take any value $\ge 0$, rather than simply 0 and 1. Now applying the
interferometer transformation
$\hat{\bm{a}}^\dagger\mapsto\bm{\Lambda}^{\rm T}\hat{\bm{a}}^\dagger$ gives
\begin{equation}
\hat\rho_{\rm trans}^{(N)} = \sum_{\bm{s}} P'_{\bm{s}} \left[ \prod_i \left(
\sum_k \Lambda_{ki} \hat a_k\dg\right)^{s_i} \ket{0}\bra{0} \prod_i \left(
\sum_k \Lambda_{ki}^* \hat a_k\right)^{s_i} \right].
\end{equation}
After detection on modes 2 to $N$, the final state obtained is
\begin{equation}
\hat\rho_{\rm out}^{(N)} = \sum_{n_1=0}^N c_{n_1} \ket{n_1}\bra{n_1}.
\end{equation}
with
\begin{equation}
c_{n_1} = \frac {K'}{n_1!} \sum_{\bm{s};\Sigma_{\bm{s}}=\Sigma_{\bm{n}}}
P'_{\bm{s}}\left| S_{\bm{s},\bm{n}}\right|^2,
\end{equation}
where $S_{\bm{s},\bm{n}} = \text{per} (\bm{\Lambda}[\bm{n},\bm{s}])$ and $K'$ is
a normalisation constant. This result is very similar to the case for inputs
with no multiphoton contribution. The only difference is that values of $s_i$
larger than 1 are now permitted, and there is the additional dividing factor of
$\prod_i s_i!$. We do not use this result in this paper, but it is a useful
general form.

Only one of the no-go theorems still applies for the case where multiphoton
inputs are allowed. For the case of a beam splitter, the multiphoton
contributions in the input give multiphoton contributions in the output.
Therefore, if zero photons are detected, there can be no improvement in the
ratio between the probability for one photon to the probability for zero
photons. In the multimode case where zero photons are detected, it is possible
to decompose the interferometer as in Fig.\ \ref{fig:line}, then omit the
U$(N-1)$ interferometer. At each beam splitter in the chain the ratio between
the probabilities for one and zero photons is not increased, so the final ratio
can not be above the maximum for the inputs. Nevertheless, the result in this
case is not as strong as in the case where the inputs have no multiphoton
contribution. Proving that the ratio between the probabilities for one and zero
photons has not increased does not prove that the absolute probability for a
single photon has not increased. The problem is that it is possible, in
principle, for the multiphoton component to be decreased sufficiently that the
probability for a single photon is increased.

If the inputs have multiphoton contributions, it is not impossible to obtain
a perfect single-photon output state. In particular, consider a state 
$\rho_0 = \sum_i q_i \ket{i}\bra{i}$, such that $q_D=0$, $q_{D+1}>0$, and
$q_i=0$ for $i>D+1$. If this state is combined with the vacuum at a beam
splitter, and $D$ photons are detected, then the output state will be a pure
single-photon state. This result also demonstrates that there is no initial
value for the single-photon probability that can not be improved upon.

The drawback to these results is that states such as $\rho_0$ would be very
difficult to produce in the laboratory. It is likely that there is some measure
of the quality of the state that can not be improved upon using linear optics
and photodetection. However, it is difficult to determine what measure this
would be. For example, it is clear that $\Pi$ can be improved, because $\rho_0$
can be heavily super-Poissonian. The same considerations rule out other simple
possibilities such as the entropy. Finding a measure that is non-increasing
under linear optics and photodetection is a promising direction for future
research.

\section{Conclusions}
\label{Sec:conc}
Triggered single-photon sources produce an incoherent mixture of zero and one
photons, with much smaller probabilities for two or more photons. Provided the
multiphoton contributions in the inputs may be ignored, we have shown that it is
possible to significantly increase the probability for a single photon by using
post-processing via linear optics and photodetection. This method has the
drawback that it produces a significant multiphoton component that is
comparable to that for the Poisson distribution.

We have shown that there are severe limitations on what post-processing
can be performed. In particular, there is an upper limit on the increase in the
probability for a single photon. This upper limit can not be achieved, but the
method we have found for increasing the probability for a single photon gives
the same scaling with the number of modes. It is likely that this method
achieves the maximum increase in the probability for a single photon. This
result is indicated numerically, but has not been proven.

In addition, it is impossible to obtain an increase in the probability for a
single photon using a single beam splitter. Alternatively, if zero photons, or
one less than the maximum input photon number are detected, it is again
impossible to obtain an improvement in the probability for a single photon. In
the restricted case that all the inputs are identical, it is impossible to
obtain an improvement in the probability for a single photon if one photon is
detected. These no-go theorems are sufficient to prove that at least a
four-mode interferometer is required to obtain an improvement.

Another important no-go theorem is that it is impossible to obtain a perfect
single-photon output with imperfect inputs. It must be emphasised that this
no-go theorem is only for mixed states with no multiphoton components. If we
relax these constraints, by considering pure superposition states of zero and
one photon, then it is possible to obtain a pure single-photon output.
Alternatively, some multiphoton states can be processed to yield a perfect
single-photon output. However, it must be emphasised that it is not likely that
pure input states, or the appropriate multiphoton states, can be produced
experimentally.

There are also a number of important unsolved problems. It is currently unknown
whether there is an upper limit (less than 1) to the initial probability for a
single photon such that it is possible to obtain an improvement. It is also
unknown if it is possible to obtain an improvement in the probability for a
single photon while maintaining zero multiphoton contribution. If such a scheme
were possible it would be very significant, because it would be possible to
obtain a state arbitrarily close to a single photon state from arbitrarily poor
input states.

\acknowledgments
RL acknowledges valuable discussions with E.\ Knill in the early stages of this
work. This work was funded in parts by the UK Engineering and Physical
Sciences Research Council. SS enjoys a Feodor-Lynen fellowship of the
Alexander~von~Humboldt foundation. RL would like to thank the NSA, NSERC and
MITACS fo support. BCS and RL would like to thank the CIAR program on
Quantum Information Processing. This research has also been
supported by an Australian Department of Education Science and Training
Innovation Access Program Grant to support collaboration in the European Fifth
Framework project QUPRODIS and by Alberta's informatics Circle of Research
Excellence (iCORE).

\end{document}